\documentclass[12pt]{article}
\usepackage[hmargin=.75in,vmargin=1in]{geometry}
\usepackage[american]{babel}
\usepackage[T1]{fontenc}
\usepackage{times}
\usepackage{array,multirow,makecell}
\setcellgapes{1pt}
\makegapedcells
\newcolumntype{R}[1]{>{\raggedleft\arraybackslash }b{#1}}
\newcolumntype{L}[1]{>{\raggedright\arraybackslash }b{#1}}
\newcolumntype{C}[1]{>{\centering\arraybackslash }b{#1}}

\usepackage{textcomp}
\usepackage{epsfig,graphicx}
\usepackage{amsfonts,amsmath,amssymb}
\usepackage{fixltx2e} 

\author{
  Amadou Moctar Kane\\
\small  \texttt{KSecurity, Dakar, Senegal. amadou1@gmail.com }
  }

\title{\bf How DNA Cryptography can help whistleblowers and refugees}           

\begin{document}

\maketitle                        

\begin {abstract}

\noindent The recent progress in DNA sequencing will probably revolutionize the world of electronic. Hence, we went from DNA sequencing that only research centers could realize, to portable, tiny and inexpensive tools. So, it is likely that in a few years these DNA sequencers will be included in our smartphones.

\noindent The purpose of this paper is to support this revolution, by using the DNA cryptography, hash functions and social networks. The first application will introduce a mutual entity authentication protocol in order to help waifs, refugees, and victims of human trafficking to find their biological parents online. 

\noindent The second application will also use the DNA cryptography and the social networks to protect whistleblowers' actions. For example, this method will allow whistleblowers to securely broadcast on social networks, their information with one grape.

\end {abstract}
\medskip
\section{ Introduction}

Inexpensive, smaller than a smartphone, and used for real-time biological analysis, the new type of DNA-sequencing MinION seems to be very performing \cite{{Camilla},{minidna}}.
The advance means that scientists in the field will no longer have to take samples from people, animals, or the environment, and send them back to a lab to have the DNA read on machines that can take days to return results. 
Camilla Ip, a co-author of \cite{Camilla, theguardian}, believes that people will soon be connecting MinIONs to smartphones, \dots ''If anyone had the ability to do DNA sequencing with a mobile phone with attachable DNA sequencer, what could you do with it?'' she said.

Hence, it is likely that Apple (to enrich HealthKit), Google (Google DNA stack), Samsung, Amazon \dots will soon engage a fierce battle to provide to their customers a DNA sequencing, in exchange of medical services tailored to their genetic characteristics. It is also certain, that this new technology will introduce some improvement in crypto applications.

We should recall here, that for security and privacy reasons, it would be very dangerous to sequence his DNA and to put it in a database own by privates companies. 

As defined in the Oxford dictionaries \cite{oxforddictionaries}, the deoxyribonucleic acid (DNA), is a self-replicating material which is present in nearly all living organisms as the main constituent of chromosomes. It is the carrier of genetic information.
The DNA is composed by molecules called nucleotides. Each nucleotide contains a phosphate group, a sugar group and a nitrogen base. The four types of nitrogen bases are adenine (A), thymine (T), guanine (G) and cytosine (C). The order of these bases is what determines DNA's instructions, or genetic code.
As defined in {Collins}, the DNA sequencing is the procedure of determining the order of base pairs in a section of DNA (deoxyribonucleic acid).
It includes any method or technology that is used to determine the order of the four bases adenine (A), guanine (G), cytosine (C), and thymine (T) in a strand of DNA.

Even though we are all unique, most of our DNA is actually identical to other people's DNA \cite{biotechlearn}. However, specific regions vary highly between people, these regions are called polymorphic. Differences in these variable regions between people are known as polymorphisms. Each of us inherits a unique combination of polymorphisms from our parents. 

As defined in \cite{dnaprofiling}, the predictable inheritance patterns is at certain locations (called loci) in the human genome, which have been found to be useful in determining identity and biological relationships. These loci contain specific DNA markers that scientists use to identify individuals. In a routine DNA paternity test, the markers used are Short Tandem Repeats (STRs).

Repetitive genetic elements are an important class of polymorphic DNA. Repetitive genetic elements include microsatellites or STRs (short tandem repeats) and the minisatellites or VNTRs (variable number of tandem repeats), which are distinguished primarily on the basis of size and repeat pattern: The repeated sequence in microsatellites range from two to six bases, while in a VNTR it ranges from eleven to sixty base pairs \cite{polymorphisms}.

In other words, short tandem repeats (or STRs) are regions of noncoding DNA that contain repeats of the same nucleotide sequence. For example, GATAGATAGATAGATAGATAGATA is a STR where the nucleotide sequence GATA is repeated six times.

Each person's DNA contains two copies of these markers, one copy inherited from the father and one from the mother. Within a population, the markers at each person's DNA location could differ in length and sometimes sequence, depending on the markers inherited from the parents.
The combination of marker sizes found in each person makes up his/her unique genetic profile. When determining the relationship between two individuals, their genetic profiles are compared to see if they share the same inheritance patterns at a statistically conclusive rate.

For example \cite{dnaprofiling}, this sample ( {\emph figure 1 }), report from the commercial DNA paternity testing laboratory Universal Genetics, signifies how relatedness between parents and child is identified on those special markers:

\begin{figure}
\begin{center}
\begin{tabular}{|R{2.5cm}|C{2.5cm}|L{2.5cm}|L{3cm}|L{2.5cm}|}
\hline \emph{DNA Marker} & Mother &  Child  & Alleged Father & Repeat motif  \\
\hline  \textbf{D21S11} & 28, 30 & 28, 31 & 29, 31 & TCTA  \\
\hline \textbf{D7S820} & 9, 10 & 10, 11 & 11, 12 & GATA  \\
\hline  \textbf{TH01} & 14, 15 & 14, 16 & 15, 16 & AATG  \\
\hline \textbf{ D13S317} & 7, 8 & 7, 9 & 8, 9 & TATC  \\
\hline  \textbf{D19S433} & 14, 16.2 & 14, 15 & 15, 17 & AAGG  \\
\hline 
\end{tabular}
\end{center}
\caption{ Relatedness between parents and child on special markers}
\end{figure}

The partial results indicate that the child and the alleged father's DNA match among these five markers. The complete test results should show this correlation on 16 markers between the child and the tested man to enable a conclusion to be drawn as to whether or not the man is the biological father.

Another example is describe in \cite{tracey}, where the basis of parentage analysis is very simple: a child must receive, in the absence of mutation, one allele matching each parent. For example, in a simple case, a mother who is a genotype 10,12 and a father who is a 9,14 may produce children of the following types: 9,10 9,12 10,14, and 12,14. If the specific genotype of a child is known, say it is 10,14, and the genotype of the parent contributing one allele, usually the mother, is also known, we may identify the allele that must have come from the other questioned parent.
In this case, the mother is known to be a 10,12; the egg must have been a 10. The true father must have contributed a sperm carrying the 14 allele. When the true father is known to be a heterozygote or an alleged father is identified as a heterozygote, he has a 50:50 chance of fathering a child with each of his alleles.

Generally each marker is assigned with a Paternity Index (PI), which is a statistical measure of how powerfully a match at a particular marker indicates paternity, but in this paper we will not work with this index.

Since the introduction of the computing DNA by Adleman \cite{adleman}, a huge study has been conducted on DNA cryptography (An overview can be find in \cite{{tornea},{jacob}}  ). There are  essentially three main area, the first one uses the DNA as  random sequence from which they build a one time pad algorithm, the second one hides and stores information in the DNA  which is a steganography  model, more recently, some plant DNA are used as watermark to fight against counterfeit electronic devices.

\textbf{Remark :}

Generally, the DNA encoded by a 4-letter alphabet: $\{A, C, G, T\}$ is transposed into binary alphabet $(A - 00, C - 01, G - 10, T - 11)$ \cite{tornea}. 

However, if we define a bijective function, $f: X \rightarrow Y$, where set $X$ is $\{1, 2, 3, 4\}$ and set $Y$ is $\{C, G, A, T\}$ (for example, $f(3) = A$), then we would find a succession of integer which is similar to the continued fraction expansion of a rational number. 

It might also be noted that the presence in the DNA of monosatellites and microsatellites (repetitive sequences) could perhaps associate each DNA with a quadratic irrational.

We recall that the continued fraction expansion of an irrationnal number is infinite, while the continued fraction expansion of a rational number is finite.

For example, if this sequence $[GCCCTCCCTCCCTCC\dots]$  is in the DNA, then with the bijection $f$, it will correspond to $[2111411141114111\dots]$ which looks like the continued fraction expansion of $\sqrt{7}$. 

We could find $\beta$ the rational number corresponding to the DNA of someone, and $\gamma$ the rational number corresponding to the DNA of somebody else, such that the DNA of all mankind will be between $ \beta $ and $ \gamma $.

There exist also, an irrational number $\alpha$ such that $\beta < \alpha < \gamma$, hence we believe that the continued fraction expansion of $\alpha$  is expected to recover much of the DNA of humans.

We recall that generalized continued fraction can be useful in stream cipher \cite{kane}.

\vspace{0.5cm}

This paper is organized as follows. In section 2, we will present a mutual authentication scheme destined to help refugee, waifs, and victims of human trafficking. The following section will present the threat model and the security given by this first scheme. The section 4 will introduce a scheme improving whistleblowers' actions, and before the conclusion we will present the security given by our last scheme.

\section{The use of DNA and social networks to find his biological parents}

The exceptional increase of the number of people separated by wars, by the human trafficking or simply by abandonment implies the obligation to find tools, which could permit victims to securely find their biological relatives. 
Our aim is to perform a mutual authentication based on the DNA while preserving the confidentiality of that latter.

As showed in section I, it is already well known that the DNA can help to find parental relations between people. 
In 2008, Bruekers et al. \cite{Bruekers} presented a paper providing an efficient and practical privacy protocol that allow the secure matching of DNA profiles. In that paper they used a homomorphic public key encryption scheme, such as Paillier encryption to design a secure identity and paternity testing. However their protocol is not secure if the participants use proper input data and our approach is different since we used the paternity testing to set up an entity authentication protocol.

Hence, in this section, we will base our work on relations (DNA polymorphic marker) already found by biologists, to build a mutual entity authentication scheme using DNA. This new scheme will also be based on a zero knowledge principle, in the sense that, we wanted to hide the DNA of the claimant while in the same time, we wanted to find his parental relation among the billions of people who are in the social networks.

The main issue in DNA authentication is the fact that someone can spoof easily some attributes, such as staling the hair of someone else. In order to avoid this issue, we use a double factor authentication.

\subsection{Definitions}

\begin{itemize}
\item  As define in \cite{menezes}, an entity authentication is the process whereby one party is assured (through acquisition of corroborative evidence) of the identity of a second party involved in a protocol, and that the second has actually participated (i.e., is active at, or immediately prior to, the time the evidence is acquired).
\item Zero-knowledge (ZK) protocols allows the prover to demonstrate the knowledge of a secret while revealing no information whatsoever (beyond what the verifier was able to deduce prior to the protocol run) of use to the verifier in conveying this demonstration of knowledge to others.
\item $H$ and $H_2$  are two different secure collision resistant one-way hash function, which takes as argument a string of arbitrary length and produces as output a binary string of a fixed length.
\item As used in \cite{wang}, we assume in this paper that hash function can insure the security of one cryptographic system with sound unpredictability.
\end{itemize}

\subsection{Mutual Authentication scheme}
As recalled in \cite{Brainard}, user authentication in computing systems traditionally depends
on three factors: something you have (e.g., a hardware token), something you are (e.g., biometrics, DNA, fingerprints or retinal scans ), and something you know (e.g., a password or a passphrase).

The two-factor authentication consists in the combined use of two different factors, among the three mentioned above.

The use of human relations is already a reality in authentication, for example in \cite{Brainard}, Brainard et al. have developed an interesting authentication mechanism using the factor ''somebody you know''. So, you can ask someone to help you connect. This new authentication method uses a voucher system, allowing for example Harry to help Alice to log in by using her password and the voucher received from Harry. 

However, in our paper, the approach that we have for human relations is different,  since we assume here that all contacts (Helpers) are trustworthy, in the same time, we make sure that the contact can not know the contents of the object that he carries, if he is not the recipient of the message.

\subsubsection{Authentication of the child}

We recall that humans have two alleles at each genetic locus, with one allele inherited from each parent. 

Let's suppose that Alice who was a victim of human trafficking (the request can also come from the biological mother or father), is trying to find her biological parents on the social networks while preserving her anonymity. 
In this two-factor authentication, she will use her DNA and something she knows, this can be her date and place of birth, the street were she grew up or something else which she shares with her biological relatives.
\newline \emph{Step 1}: Starting from the 16 well known polymorphic markers (D8S1179, D21S11, D7S820, CSF1PO, D3S1358, TH01, D13S317, D16S539, D2S1338, D19S433, \dots) she will concatenate each of her alleles with her date of birth.
\\ \emph{Step  2}: She will hash the result of each concatenation with the hash function $ H $.
\newline \emph{Step  3}: She will take one character (hexadecimal) on each hash obtained, e.g. she could take the last element of the hash.
\newline \emph{Step  4}: She will send the sets (obtain from the two alleles of each marker) of characters taken previously to all of her contacts who are on social networks.
\newline \emph{Step  5}: Those who will receive this request and who are not concerned by any these social problems (abandonment, victim of traffic, refuges ...) will simply relay the information to all their friends who are on social networks.
\newline \emph{Step  6}: Those who are concerned by these social problems (abandonment, traffic, refuges ...) will repeat the previous \emph{steps 1-3}: concatenate the birth date of the missing person with the different alleles of his own DNA, hash the concatenation and create the sets. 
\newline \emph{Step 7}: Those who are concerned by these social problems (abandonment, traffic, refuges \dots) will compare sets obtained in \emph{Step 6} with those received in \emph{Step 5}. 

If for each polymorphic marker, there exist at least one character of the set sent by the prover corresponding to one element of the set created by the verifier, then the verifier will consider that the applicant knows the DNA and the date of birth of the person sought.

The probability of having on each of the 16 markers a correspondance due to chance is very low $(\frac{1}{8}) ^ {16}$.
Hence, having 16 matching will allow the parent to know that there's a high probability, that the sender is his child who has been victim of human trafficking.

If he has a reasonable number of matching (15 on the 16 markers), then he may think that it is also his child and the difference may be due to a possible mutation.

\subsubsection{Authentication of the parent}

The parent (Bob) who received the request, will now prove to his child, who was a victim of human trafficking, that he is his biological parent with the following steps.
\newline \emph{Step  8}: He will concatenate each of his alleles with the birth date of the missing child.
\newline \emph{Step  9}: He will hash the result of each concatenation, with the hash function $H_2$ (a hash function other than that used at the beginning of the algorithm).
\newline \emph{Step  10}: He will take a portion of each hash obtained, e.g. he could take the last character of the hash ( in hexadecimal).
\newline \emph{Step  11}: He will send the sets of two elements taken from the hash to all his contacts, on social networks.
\newline \emph{Step  12}: Those who will receive this request and who are not concerned by these social problems (abandonment, victim of traffic, refuges ...) will simply relay it to all their friends, who are on social networks.
\newline \emph{Step  13}: Those who are concerned by these social problems (abandonment, victim of traffic, refuges ...) will repeat the previous \emph{steps 8-10}: concatenate the date of birth of the missing with the different alleles of his own DNA, hash the concatenation previously obtained, and form the sets of two elements.
\newline \emph{Step  14}: Those who are concerned by these social problems (abandonment, victim of traffic, refuges ...) will compare sets obtained in \emph{Step 13} with those received in \emph{Step 12}. 

If he has 16 on 16, then he can say that he found his biological parent, otherwise if he has a reasonable number of matching (15 on the 16 markers), then he may think that it is his biological parent and the difference may be due to a possible mutation.

\subsubsection{The communication between the parent and the child}

When the parent and the child have identify alleles  which seems to be shared between them, they can  form a  secret key with these DNA polymorphic markers, by concatenating these markers,  adding a random padding and hashing the all with a good hash function for example \emph{\emph{SHA2}512}.

The Child/Parent will send the encrypted message to all his contacts in social media, these latter will send it to their contacts, until it reaches the real owner, who will be able to open the message since he also has the secret key.

\medskip
\textbf{Remark}

\begin{itemize}
\item  This method can also be applied to monozygotic twins.

\item This algorithm can be used for someone who has to prove his innocence and who does not wish to submit his DNA to the judicial authorities, this one could indeed work on the bits obtained from the hashed of his DNA, without revealing this latter. In other words, the judicial authorities could use a zero knowledge principle (via a hash function), without revealing people's DNA.

\item This algorithm can also be used by banks, for the mutual authentication of their clients.

\item  This method will also allow to great specialists, who are also present on social networks, to diagnose genetic diseases without knowing the identity of the applicant.

\item  Unfortunately, we also may note that people with bad intentions (secret service, pranksters, \dots) could use this method, in their job.

\item  The date of birth that we use here is just an example, one could use other authentication factors, such as the place where the child has been abandoned, or the day of his disappearance, or the name of his best friend before  his disappearance etc.

\item The random padding works as follows: let's suppose that all the users ( including the parent and the child) have a machine (it can be a token or a software), which would produce a pseudo random padding, depending on the date when you enter your demand and depending on the second factor authentication, for example, here it can be the place and birth date of the child.

\end{itemize}

\subsubsection{Example}
In this example, we will work with this five polymorphic markers used in \emph{figure 1} : \emph{D21S11 , D7S820, TH01, D13S317, D19S433}. 

Mr. Rousseau wants to find his child, that he had abandoned a few years ago to the public assistance.

We will suppose that the child is born on 1/1/1747.

{\bf Preliminaries}

Let's suppose that the DNA of Mr. Rousseau can be presented as follows:

\emph{Marker D21S11}

29, 31 with the repeated motif TCTA.

Allele1 (29): TCTATCTATCTATCTATCTATCTATCTATCTATCTATCTATCTATCTATCTATCTATCTA TCTATCTATCTATCTATCTATCTATCTATCTATCTATCTATCTATCTATCTATCTA.

Allele2 (31): TCTATCTATCTATCTATCTATCTATCTATCTATCTATCTATCTATCTATCTATCTATCTA TCTATCTATCTATCTATCTATCTATCTATCTATCTATCTATCTATCTATCTATCTATCTATCTA.

\emph{Marker D7S820}

11, 12 with the repeated motif GATA

Allele1 (11): GATAGATAGATAGATAGATAGATAGATAGATAGATAGATAGATA.

Allele2 (12): GATAGATAGATAGATAGATAGATAGATAGATAGATAGATAGATAGATA.

\emph{Marker TH01}

15, 16 with the repeated motif AATG

Allele1 (15): AATGAATGAATGAATGAATGAATGAATGAATGAATGAATGAATGAATGAATGAATG AATG.

Allele2 (16): AATGAATGAATGAATGAATGAATGAATGAATGAATGAATGAATGAATGAATGAATG AATGAATG.

\emph{Marker D13S317}: 8, 9 with the repeated motif TATC

Allele1 (8): TATCTATCTATCTATCTATCTATCTATCTATC.

Allele2 (9): TATCTATCTATCTATCTATCTATCTATCTATCTATC.

\emph{Marker D19S433}

15, 17 with the repeated motif AAGG

Allele1 (15): AAGGAAGGAAGGAAGGAAGGAAGGAAGGAAGGAAGGAAGGAAGGAAGGAAGG AAGGAAGG.

Allele2 (17): AAGGAAGGAAGGAAGGAAGGAAGGAAGGAAGGAAGGAAGGAAGGAAGGAAGG AAGGAAGGAAGGAAGG.

Let's suppose now that the DNA of Rousseau's child can be presented as follows:

\emph{Marker D21S11}: 28,31 with the repeated motif TCTA

\emph{Marker D7S820}:  10,11 with the repeated motif GATA

\emph{Marker TH01}: 14,16 with the repeated motif AATG

\emph{Marker D13S317}: 7, 9 with the repeated motif TATC

\emph{Marker D19S433}: 14, 15 with the repeated motif AAGG.

\medskip

{\bf Steps}

\smallskip

{\bf Step 1, 2, 3:} 

\smallskip

\emph{Marker D21S11}

\emph{SHA1}(TCTATCTATCTATCTATCTATCTATCTATCTATCTATCTATCTATCTATCTATCTATCTATCTATCTA TCTATCTATCTATCTATCTATCTATCTATCTATCTATCTATCTATCTA1/1/1747)= 1f 8e 0f 2c fb a1 5c 75 b9 be dc 0d 5c e7 4f 3a d0 f2 63 e1

\emph{SHA1}(TCTATCTATCTATCTATCTATCTATCTATCTATCTATCTATCTATCTATCTATCTATCTATCTATCTA TCTATCTATCTATCTATCTATCTATCTATCTATCTATCTATCTATCTATCTATCTA1/1/1747)= a8 d9 31 13 5e 14 20 2c 5a 26 c5 80 41 9e ca 9c 3d 40 2d 6a

The set which Mr. Rousseau should includ in the request: \emph{(1, a)}.

\emph{Marker D7S820}

\emph{SHA1}(GATAGATAGATAGATAGATAGATAGATAGATAGATAGATAGATA1/1/1747)= 7f 3d ff 9d 45 ca 9d 88 e7 e4 18 fe 97 ad 48 88 8a 80 4d 7b

\emph{SHA1}(GATAGATAGATAGATAGATAGATAGATAGATAGATAGATAGATAGATA1/1/1747)= d0 32 a3 2d 66 3c 1d eb 82 1a 61 6a 8d e8 09 06 e5 ea b1 04

The set which Mr. Rousseau should includ in the request: \emph{(b,4)}.

\emph{Marker TH01}

\emph{SHA1}(AATGAATGAATGAATGAATGAATGAATGAATGAATGAATGAATGAATGAATGAATGAATG 1/1/1747)= 9e 70 78 3b e2 67 cc 3f 93 76 77 db 1e 59 ba 4d d0 30 58 13

\emph{SHA1}(AATGAATGAATGAATGAATGAATGAATGAATGAATGAATGAATGAATGAATGAATGAATG AATG1/1/1747)= 2a e5 d6 f8 f4 40 da 24 3d 56 54 67 f3 6d cd 43 5a 86 05 d7

The set which Mr. Rousseau should includ in the request: \emph{(3,7)}.

\emph{Marker D13S317}

\emph{SHA1}(TATCTATCTATCTATCTATCTATCTATCTATC1/1/1747)= 38 76 fc a6 55 d5 d8 42 78 c3 84 00 ac dc 5c 5e bb 64 90 75

\emph{SHA1}(TATCTATCTATCTATCTATCTATCTATCTATCTATC1/1/1747)= 1c 55 80 16 89 b6 2f a5 39 0e 26 33 8f 12 6f 93 89 02 8d 59

The set which Mr. Rousseau should includ in the request: \emph{(5,9)}.

\emph{Marker D19S433}

\emph{SHA1}(AAGGAAGGAAGGAAGGAAGGAAGGAAGGAAGGAAGGAAGGAAGGAAGGAAGGAAGG AAGG1/1/1747)= f6 46 f7 88 3d 31 50 22 33 a5 a1 cd 81 d8 75 e7 0d ba ad b8

\emph{SHA1}(AAGGAAGGAAGGAAGGAAGGAAGGAAGGAAGGAAGGAAGGAAGGAAGGAAGGAAGG AAGGAAGGAAGG1/1/1747)= f5 27 83 e6 1d ef a8 34 a8 7e b0 91 42 43 f7 1a bc ae 28 35

The set which Mr. Rousseau should includ in the request: \emph{(8,5)}.

{\bf Step 4:}

Mr. Rousseau will send to all of his contacts on social networks, including Mr. Diderot, the following sets and the name of the hash function used. 

\{Marker D21S11 (1, a); Marker D7S820 (b,4); Marker TH01 (3,7); Marker D13S317 (5,9); Marker D19S433  (8,5). \emph{SHA1}\}.

{\bf Step 5:}

Mr. Diderot who is not looking for a missing child, sends the packages to all his friends, who are on social networks, including Mr. Voltaire.
Mr. Voltaire who is not interested by any search of an abandoned child, will simply convey the message to his friends who are on social networks including Mr. Mouchaard.

{\bf Step 6, 7: }

Let's suppose that Mr. Mouchaard is looking for a child that he had abandoned and who is born on 02.25.1749 (2/25/1749). The DNA of Mr. Mouchaard is presented as follows:

\emph{Marker D21S11}: 23, 25 with the repeated motif TCTA

Allele1 (23): \emph{SHA1}(TCTA \dots TCTA2/25/1749)= \dots 6

Allele2 (25): \emph{SHA1}(TCTA \dots TCTA2/25/1749)= \dots 6

The set created is (6, 6).

\emph{Marker D7S820}: 6, 11 with the repeated motif GATA

Allele1 (6): \emph{SHA1}(GATA \dots GATA25/2/1749)= \dots1

Allele2 (11): \emph{SHA1}(GATA \dots GATA25/2/1749)= \dots8

The set created is (1, 8).

\emph{Marker TH01}: 10, 12 with the repeated motif AATG

Allele1 (10): \emph{SHA1}(AATG \dots AATG2/25/1749)=\dots 5

Allele2 (12): \emph{SHA1}(AATG \dots AATG2/25/1749)=\dots 9.

The set created is (5, 9).

He can stop the calculations, since none of the first three sets he found did match to sets that are in the request. Thus, for Marker D21S11 instead of (1, a), he has obtained (6, 6); for Marker  D7S820 instead of (b, 4), he has obtained (1, 8); for Marker TH01 instead of (3, 7), he has obtained (5, 9).

Mr. Mouchaard can conclude, that it is not his child, hence, he can send the message to his contacts who are on social networks, including Rousseau's child.

By following the same procedure as Mr. Mouchaard, Rousseau's child will get this result:

\{Marker D21S11 (3,a) ; Marker D7S820 (5, b); Marker TH01 (e, 7); Marker D13S317 (9, 9); Marker D19S433 (8, 5). \emph{SHA1}\}.

In comparison, for Marker D21S11 (1, \textbf{a}) sent by Mr. Rousseau, he obtained (3,\textbf{a}); for Marker D7S820 (\textbf{b},4), he obtained  (5, \textbf{b}); for Marker TH01 (3,\textbf{7}),  he obtained (e, \textbf{7}); for Marker D13S317 (5,\textbf{9}), he obtained (\textbf{9}, \textbf{9}); for Marker D19S433 (\textbf{8},5) he obtained (9, \textbf{8}).

Rousseau's child is confident that the request comes from someone who knows his DNA or the DNA of one of his parents. Also, the author of the request knows the second authentication factor (here it is the date of birth).
He can reasonably think, that it is one of his parents
(With the presence of Y-STR or mitochondrial DNA, he could know whether the DNA comes from his father or mother).

\smallskip

{\bf Step 8, 9, 10:}

Now, in order to authenticate himself, Rousseau's child will use his DNA as follows:

\emph{Marker D21S11}: 28,31 with the repeated motif TCTA.

Allele1 (28): \emph{SHA2}(TCTA \dots TCTA1/1/1747)= \dots 0d

Allele2 (31): \emph{SHA2}(TCTA \dots TCTA1/1/1747)=\dots 02

The set which Rousseau's child should includ in his request is: (d,2).

\emph{Marker D7S820}: 10,11 with the repeated motif GATA.

Allele1 (10): \emph{SHA2}(GATA \dots GATA1/1/1747)=\dots 96

Allele2 (11): \emph{SHA2}(GATA \dots GATA1/1/1747)=\dots 1a

The set which Rousseau's child should includ in his request is: (6,a).

\emph{Marker TH01}: 14,16 with the repeated motif AATG. 

Allele1 (14): \emph{SHA2}(AATG \dots AATG1/1/1747)= \dots f3

Allele2 (16): \emph{SHA2}(AATG \dots AATG1/1/1747)= \dots fc

The set which Rousseau's child should includ in his request is: (3,c).

\emph{Marker D13S317}: 7, 9 with the repeated motif TATC. 

Allele1 (7): \emph{SHA2}(TATC \dots TATC1/1/1747)=\dots a1

Allele2 (9): \emph{SHA2}(TATC \dots TATC1/1/1747)=\dots 43

The set which Rousseau's child should includ in his request is: (1,3).

\emph{Marker D19S433}: 14, 15 with the repeated motif AAGG.

Allele1 (14): \emph{SHA2}(AAGG \dots AAGG1/1/1747)=\dots 5F

Allele2 (15): \emph{SHA2}(AAGG \dots AAGG1/1/1747)=\dots B1

The set which Rousseau's child should includ in his request is: (F,1).

Rousseau's child will send to all of his contacts on the social networks \{Marker D21S11 (d,2); Marker D7S820 (6,a);  Marker TH01 (3,c); Marker D13S317 (1,3);  Marker D19S433 (F,1); \emph{SHA2}\}.

We recall that an attacker can not distinguish, those who issue the requests, from those who relay requests.

{\bf Step 11, 12, 13, 14}

Mr. Mouchaard will verify if this request matches with his DNA and the birth date of his abandoned child.

\emph{Marker D21S11}: 23, 25 with the repeated motif TCTA.

\emph{SHA2}(TCTA \dots TCTA25/2/1749)= \dots 6F

\emph{SHA2}(TCTA \dots TCTA25/2/1749)= \dots4F  

set obtained (f,f).

\emph{Marker D7S820}: 6, 11 with the repeated motif GATA.

\emph{SHA1}(GATA \dots GATA25/2/1749)= dots 1A 

\emph{SHA1}(GATA \dots GATA25/2/1749)= \dots 85

set obtained (a,5).

\emph{Marker TH01}: 10, 12 with the repeated motif AATG.

\emph{SHA1}(AATG \dots AATG25/2/1749)= \dots 6E 

\emph{SHA1}(AATG \dots AATG25/2/1749)= \dots 88 

set obtained (e,8).

\emph{Marker D13S317 }: 7, 6 with the repeated motif TATC.

\emph{SHA1}(TATC \dots TATC25/2/1749)= \dots 99 

\emph{SHA1}(TATC \dots TATC25/2/1749)= \dots 18 

set obtained (9,8).

He will obtain the following sets: Marker D21S11: (f,f) Marker D7S820: (a,5) Marker TH01 (e,8) Marker D13S317 : (9,8). 

He does not need to finish the list of the markers since, the following comparison {Marker D21S11 (d, 2) (sent by Rousseau's child) with (f, f); Marker D7S820 (6 a) with (a, 5); Marker TH01 (3, c) with (e, 8); Marker D13S317 (1.3) with (9.8)} shows that this person is not the biological child of Mr. Mouchaard.

Mr. mouchaard will then send the request, to all his friends present on social networks including Mr. Voltaire.

{\bf Step 11, 12, 13, 14}

Mr. Voltaire will send it to all his friends who are on social networks (including Mr. Diderot)

Mr. Diderot will send the message to all his contacts who are online (including Mr. Rousseau).

{\bf Step 11, 12, 13, 14}

Mr. Rousseau will verify that the one who claims to be his child is right.

\emph{Marker D21S11}: 29, 31 with the repeated motif TCTA

\emph{SHA2}(TCTA \dots TCTA1/1/1747)= \dots 2A

\emph{SHA2}(TCTA \dots TCTA1/1/1747)= \dots 02

set obtained (A,2 )

\emph{Marker D7S820}: 11, 12 with the repeated motif GATA

\emph{SHA2}(GATA \dots GATA1/1/1747)= \dots 1A
 
\emph{SHA2}(GATA \dots GATA1/1/1747)= \dots 0C

set obtained (A,C)

\emph{Marker TH01}: 15, 16 with the repeated motif AATG

\emph{SHA2}(AATG \dots AATG1/1/1747)= \dots 6A

\emph{SHA2}(AATG \dots AATG1/1/1747)=\dots FC

set obtained (A,C)

\emph{Marker D13S317}: 8, 9 with the repeated motif TATC 

\emph{SHA2}(TATC \dots TATC1/1/1747)= \dots C1 

\emph{SHA2}(TATC \dots TATC1/1/1747)= \dots 43

set obtained (1,3)

\emph{Marker D19S433} 15, 17 with the repeated motif AAGG

\emph{SHA2}(AAGG \dots AAGG1/1/1747)= \dots B1

\emph{SHA2}(AAGG \dots AAGG1/1/1747)= \dots 07 

set obtained (1,7).

Rousseau's child has sent to all of his friends who are in social networks \{Marker D21S11 (d,2); Marker D7S820 (6,a);  Marker TH01 (3,c); Marker D13S317 (1,3);  Marker D19S433 (F,1). \emph{SHA2}\}.

Rousseau will compare \{ Marker D21S11 (D, {\bf 2}) (sent by Rousseau's child) with (A, {\bf 2} ); Marker D7S820 (6, {\bf A})  with ({\bf A}, C); Marker TH01 (3, {\bf C}) with (A, {\bf C}); Marker D13S317 ({\bf 1}, {\bf 3}) with ({\bf 1}, {\bf 3}); Marker D19S433 (F, {\bf 1}) with ({\bf 1}, 7). \emph{SHA2}\}.

Rousseau and his child have authenticate themselves, however, in order to send a secret message using their DNA, they should add another round in order to know the allele which they share in \emph{Marker D13S317}.

From the previous work, Mr. Rousseau knows that he probably shares with his child the following alleles:
in Marker D21S11, 31 with the repeated motif TCTA; in Marker D7S820, 11 with the repeated motif GATA; in Marker TH01, 16 with the repeated motif AATG; and in Marker D19S433, 15 with the repeated motif AAGG. 
However for Marker D13S317 he ignores which allele he shares with his child ( it can be the two alleles or one allele).
 
Hence, Rousseau will send a new message using another hash function \emph{SHA2-384}, to all his correspondents (in social networks) who do not know if it is a new message, who is the authors, etc. 

\{Marker D21S11 (9, D), Marker D7S820 (5,A), Marker TH01 (A,1), Marker D13S317 (9,F), Marker D19S433 (2,7) \emph{SHA2-384}\}.

When this message arrives to Rousseau's child, he will compute these Markers using the indicated hash function, and he will obtain: \{ Marker D21S11 (5, D); Marker D7S820 (F,5); Marker TH01 (B,1); Marker D13S317 (0,F); Marker D19S433 (3, 2)\}.

In Marker D13S317 we have (9, \textbf{F}) and (0, \textbf{F}) which means that they have the same allele frequency which is  9 with the repeated motif TATC.  

The secret key $K$ will look like this:
\emph{SHA2} 512(TCTATCTATCTATCTATCTATCTATCTATCTATCTATCTATCTATCTATCTATCTATCTATCTATCTA TCTATCTATCTATCTATCTATCTATCTATCTATCTATCTATCTATCTATCTATCTAGATAGATAGATAGATAGATA GATAGATAGATAGATAGATAGATAAATGAATGAATGAATGAATGAATGAATGAATGAATGAATGAATGAAT GAATGAATGAATGAATGTATCTATCTATCTATCTATCTATCTATCTATCTATCAAGGAAGGAAGGAAGGAA  GGAAGGAAGGAAGGAAGGAAGGAAGGAAGGAAGGAAGGAAGG1/1/1747atparishospital).

{\bf Communication}

With $K$, Rousseau's child will encrypt his message and send the cipher text to all his contacts. Those who already have a secret key will try to decrypt the message, otherwise, they will simply relay the encrypted message. 
Hence, here, Mr. Mouchaard, Mr. Voltaire, and Mr. Diderot will simply relay the encrypted message, while Mr. Rousseau will be able to read the message.

\section{Security}
\subsection{Threat Model}
Our first objective is to allow people to look for their biological parents anonymously on social networks.
Our scheme is based on the three following elements: 
The users, the communication channels and the authentication factors

{\bf Users:}

The parties involved in the protocol are the missing person, his biological parents (father and / or mother), and all persons present on social networks, including the adversaries. Therefore there exists many groups of participants:

\begin{itemize}
\item  The set of all those who are looking for their parents or their children.
\item The set of those who are helping people to find their parents or their children.
\end{itemize}
The intersection of these two sets is not empty, since among those who are helping in the search, some are also looking for their parents.

Among the participants we also have adversaries who are:
\begin{itemize}
\item  Those who are trying to prevent the reunion between parents and children (destroy or modify the requets).
\item Those who are trying to impersonate the biological parents or children for a malicious reason (inheritance, etc.).
\item Those who are trying to find the owners of the requests (remove the anonymity of parents and children).
\end{itemize}

{\bf Communication channels:}

Even if the communication channels used can be considered as secure, since some social networks use the end-to-end encryption, we consider here that the communication channel that we use is partly secure. Due to the fact that an adversary can read a message destined to a potential target.

{\bf Authentication factors}

We believe that parents and children who have been separated could share common elements such as the date or place of birth of the child. By cons, the name of the mother or father can be ignored by the child. We define by $ S $ the set of memories shared by parents and children. $ S \neq \emptyset$, as if it is empty the authentication would be based on one single factor, which is the first factor (DNA).

We believe that the DNA can be spoofed (for example saliva and hair are easy to obtain \dots) by an active opponent.

That is to say, the second factor of authentication must be difficult to guess even for a close parent, for example, instead of ''life questions" which can focus on "What is the name of your first pet?" this factor must be primarily based on family secrets.

Hence, we consider in this scheme, that the probability for an adversary to be able to guess the second authentication  factor is close to zero.
\subsection{Threat in our scope}

As a possible threat, we consider that an adversary can collude with other users in order to set up an attack.
We also assume that a social engineering can be mounted by an adversary.

We include in our model, an adversary who can not have a physical access to the DNA of the biological parents although he knows the child, the biological parent and the birth date of the child (for example, hundreds of babies of parents killed by Argentina's military dictatorship were given new families and identities, their true origins kept a secret, or in France, hundreds of children were stolen from reunion's island territory in official scheme to boost French rural population in 1960s and 70s \cite{argentina, france}).

We take into account an adversary who is trying to uncover the identities of biological relatives.
The adversary can also store requests sent on the social networks, and in this regard, he can specially target one social networks' user.

Finally, the threat includes an adversary who can try a flooding attack or a denial of service.

\subsection{Threat not in our scope}

We exclude in our scheme an adversary (a family member) who can guess the two authentication factors. 

We exclude from our scheme, an opponent who has the ability to close social networks and forbit emails or text messaging  in a territory, such as a government which decides to close social networks.

We also exclude a collusion of all the user's contacts against him (at least one would refuse to collude against him).

The adversary has no control over users' smartphones, hence we exclude any form of attack using spyware or malware in our scheme, since this protection should depend on each user.

\subsection{Security-Related Assumptions}

The random padding introduced can be a software or a token that users will use in their authentication.
We assume that the software or the token used is secure, i.e. an opponent can not compromise it.

\subsection{Security properties}

\begin{itemize}

\item Brute force attack against the secret key: The first protection is the mutation which could occur between the parent and the child, since the attacker will not be able to predict the exact location of this mutation.
 
Let's suppose now, that in each marker the allele frequency can take on average 10 different values, the age of the user varies from 18 to 100 and there at least 1000 hospitals were the child can born, then in order to find the correct key, the attacker should perform ${10}^{16}*365*93*1000$ tests. Hence, even if the security of the key is also based on the second factor authentication (''something you know''), it should be interesting to use more polymorphic markers (12 seems to be the frequent number currently and 100 can be a correct number against a brute force attack).

\item Impersonation attack:  in this scheme, the attacker can find a component allowing to retrieve the DNA before using social engineering to try to retreive the second authentication factor.
 To prevent this attack we could combine multiple authentication factors (''something you know''), as the place where the child was abandoned, the day of his abandonment, the location where the child would have a blotch on the skin etc.

Another solution would be to use the authentication method ''somebody you know'', introduced by \cite{Brainard}, it will allow us to exclude people who are not trustworthy, in the chain of people who forward requests. Hence, a contact who has no voucher (because a contact can know that his friend is a scam specialist) can not forward a message or when he transfers messages it will be with warnings.

However, the best protection against social engineering is to keep some secrets for himself.

\item Government spying:
A government may have the resources which permit it to register all requests that pass through networks.
Thus, by specifically targeting a user it might know all that this latter has received and sent during a period of time. However, the end-to-end encryption, and the indistinguishability of the cipher allow the user to hide his queries (e. g. he might replace the request received from one of his contacts by his own request).

In order to be sure that the request comes from a target person, the government must have a control over all contacts of the targeted person.

In order to prevent the spread of the targeted user's requests, the government must stop all the requests sent by the targeted user.

\item Collusive attack: A group of social networks' users may collude in order to break the anonymity of a parent or a child. However this attack should fail, due to the fact that the attacker must have the control of all parents' (or child's) contacts, which is difficult to achieve, since we consider that all the contact are protected against spyware and he can not collude with all of them. 

\item Attack on the anonymity: 

\begin{itemize}

\item Let's suppose that Eve tries to verify if Alice has abandoned one of her child, she would need the DNA of Alice, which could be easy to obtain (hair, saliva, etc.), but she will also need to know the second factor of the authentication, (what Alice knows) which is here for example the birth date of that supposed child.

\item Let's suppose that Alice sends a request to Eve. Eve has a doubt on the identity of the sender, she believes that it can be Alice's request and not a relay ( people must not be able to  make a difference between the owner of the request and the one who simply relays the message). Even if Eve knows the DNA of Alice we believe that she can not find the link between Alice and her child, if she does not have the second factor authentication.

\item An attack on the request can not allow someone to find the DNA of the sender, since the information is obtained from a one way hash function. In addition, there exist a large number of  input (concatenation of DNA and second factor authentication) which can produce the same output. For instance in {\bf \emph{section 2.2.4}}, \emph{Marker D13S317} produces the same output (we take the last character as an output) while their input is different.

\emph {Marker D13S317}

\emph{SHA2}(TATCTATCTATCTATCTATCTATCTATC1/1/1747) 6c b8 f0 a5 57 3a 46 9d 6e 6a 40 d5 7a b6 2e de 12 3b a5 bd 34 8f de fa d9 99 3a ee 1e c7 8c a\textbf{1}

\emph{SHA2}(TATCTATCTATCTATCTATCTATCTATCTATC1/1/1747)= CF 0A D8 01 DC 49 45 AB 8D E8 EC 30 ED 73 E5 F8 E4 15 89 6F 08 D5 3E 76 86 02 4E 6C 5A AA 12 C\textbf{1}.

\item  An algebraic attack can be attempted by someone who monitors the networks, however, the attack will be difficult to realise due to the fact that the hash values are secret, and only two characters of that hash are sent on the social network. Another difficulty is the fact that one can not distinguish Rousseau requests from those of Diderot, Paul, John \dots

\end{itemize}
\item Denial of service: An attacker may try to destoy or to modify any message receive from a targeted parent (or child) in order to prevent a reunion with his missing child, however this attack will fail since one contact is suffisant to spread the inquiry notice.
\item Flooding attack: This attack can be tried by a powerful adversary or a gouvernement, however it will fail if the number of request is limited for each user during a period of time.

\end{itemize}

\section{A toolkit for whistleblowers}

There exist some tools which are used \cite{Dingledine} or have been developed \cite{roth} for  whistleblowers. However using these tools seems to be very difficult, for example, let's suppose that the whistleblower Alice is working for a corrupted government. She wants to denounce that government to journalists or to activists such as Bob.
Today, Alice has to find and to talk with journalists or activists online, which is a very big risk, since most of journalists or activists can be under surveillance. 

Other solutions, such as the use of Tor is sufficient to create suspicion and surveillance for whistleblowers.

 In the same time, using the ingenious AdLeaks \cite{roth} when the aggregators are dishonest (which can be easy for some governments) can be dangerous for whistleblowers.

The same problem remains if the whistleblowers try to meet the journalist physically. 

In order to avoid the problems listed above, one solution would be to use the DNA cryptography.

\subsection{Our solution}

\subsubsection{Preliminaries}
Let's recall that the Japanese flower Paris japonica, has the largest genome described, meaning nearly 150 billion base pairs, i. e. 50 times the size of the human genome. This large genome means nearly ${4}^{150 billion}$ combinations, which is a huge possibilities for encryption and steganography. 

As recalled by \cite{tornea}, random numbers can be generated from DNA sequences which can be found in genetic databases in digital form.

Similarly,  other studies have shown that DNA sequences have an interesting link with random or pseudo-random sequence \cite{Gearheart, Babatunde}.

By not having the study of the DNA sequences of all the existing organisms in the universe, we will consider that the DNA sequences of an organism taken randomly produces a random sequence.

As used in \cite{wang}, we will also accept  that, some hash function (namely SHA2- 512) can insure the unpredictability of produced random number (extracted from the DNA sequence).

In order to send the secret to journalists, the first idea would be to insert the secret message in the bacteria's DNA using the steganography method presented in \cite{Palacios}.     

Unfortunately, the use of this interesting method \cite{Palacios} needs some special materials (fluorescent bacteria, nitrocellulose, transport of bacteria, \dots) which implies that this first idea is actually unrealistic, for all the potential whistleblowers. Therefore we should replace it by realistic schemes, using tools which could be available in the near future.

\subsubsection{DNA cryptography for whistleblower}
Let's suppose that the whistleblower can buy any foods, flowers or books without creating suspicions on his activities. He can also send messages to his contacts or followers on the social networks without creating suspicions.

{\bf The whistleblower's actions}

The steps which can permit the whistleblower to reveal information without any possibility of the government's retaliation will look like a ''message in a bottle'' thrown in the sea:

\emph{Step 1}: Alice will buy randomly at least ten different organisms (it can be a flower, a crab, a hot-dog, a potato, a melon, a fish, a rabbit, etc.) and ten different used books (a bestseller can be a good choice). 
\\ \emph{Step 2}: She will sequence all the organisms (even partly, since she just needs a part of their DNA), and extract from each DNA its first 1000 based-pair. 
\\ \emph{Step 3}: For each book purchased, she will open randomly the book, take one sentence on each page opened (it should be defined in the standard, for example, it can be the first sentence) and insert in that opened page, a part of the organism bought. 
\\ \emph{Step 4}: She will choose randomly ten different addresses in the country and she will deposit or ''forget'' discreetly in each of these places one book.
\\ \emph{Step 5}: She concatenates the extracted part of each DNA with the sentence taken in the book and the exact address where the corresponding book has been ''forgotten''.
\\ \emph{Step 6}: She  hashes the result of the previous concatenation with a good hash function (to date, it is well accepted that \emph{SHA2-512 } is good hash function), and use the hash obtained as a symmetric key.\\ \emph{Step 7}: She encrypt the file (proof of corruption) with the symmetric key and sends the cipher to all of her contacts, on the social networks using the end-to-end encryption principle (for example Whatsapp, Minds \dots). Hence, contacts will receive among their normal messages one which seems to be encrypted. 

{\bf The Helper's actions}

What should do, the Helper who picks up in the street an organism or a book containing a particular organism (fish, fruit, vegetable, steak, flower, ...)? \\ \emph{Step 9}: The Helper should sequence this organisms, and extract from their DNA the first 1000 based-pair.\\ \emph{Step 10}: He creates his symmetric key by concatenating the extracted part of the DNA with the exact address where he has found the book and the sentence taken on the page.\\ \emph{Step 11}: Whenever he receives an encrypted message from his contacts, he will verify if the symmetric key can decrypt that message.\\  \emph{Step 12}: when he receives and decrypts the file $X$ containing the proof of corruption, he should first wait for a while (in order to hide the origin of the message), before publishing it or sending it to an activist or a journalist.

\textbf{Remark}

\begin{itemize}

\item In order to clean the network, any emitted message should have an end of life. 

\item The cost of this operation is about two thousand dollars, since the cost of the mini DNA sequencer is about one thousand dollars.

\item If someone gets caught with a usb key, containing secret information or secret keys in an airport, he may face reprisals while the fact of getting caught with an apple presents no risk. In the same time, people may be reluctant to insert an usb key found somewhere in the street (virus, spyware, \dots), so the use of mango, or raisin, or banana  can be the best alternative.

\item Using a book is optional, as we just need to highlight a organism in the street.

\item As soon as he creates his symmetric key, he should destroy the evidences  (book, fruits, etc).

\item It is likely that an individual, who is in a country where freedom of speech is limited and who wishes to send evidence of corruption, to journalists based in an other continent, would send for example, seasonal fruits to Helpers, who would relay it to journalists.

\item The fact that people accept to transfer encrypted messages to their contacts, on social networks, and the sequencing of organisms found in streets, can increase the chances of success of whistleblowers.

\item If we suppose that every contact has at least ten correspondents (followers, friends, \dots), we will have at the hundredth relay (rettiwt, \dots), around ${10}^{100}$ viewers (some will receive the message a several time). 

\item The whistleblower should hide his message among other messages. 
For example, when he receives an encrypted message from one of his contact, instead of relaying this one, he will delate it and relay his own message.

\end{itemize}

\subsection{Example}

Mr. Kent who is working for a corrupted government will send to his son, who lives in another continent a bunch of grapes. 

Let's suppose that he has a file $X$ showing that his government is corrupted. 

He will create a symmetric key by hashing the 1000 first based-pair of the raisin's DNA. 

He will encrypt $X$ with the symmetric key, and insert the cipher, with a good steganography software, on the family movie published on Facebook. 

When his son receive the grape, he extract from the family movie (with the public steganography method) the encrypted file. He will also sequence the DNA of the raisin and hash the first thousand characters of the raisin DNA. 

Hence, he will find $X$ and give that file to journalist or activist all around the world, since he is not under surveillance.

\section {Security}
\subsection{Threat Model}
As supposed in \cite{roth}, we also consider in our scheme that the primary security objective of this tool is to conceal the presence of whistleblowers, and to eliminate network traces that may make one suspect more likely than another in a search for a whistleblower. 

We consider in this scheme that the security of the whistleblower is more important than the disclosure of the information.

Here, our scheme will be based on the following elements: the whistleblower, the communication channels, the Helpers and the adversaries.

{\bf The whistleblower}

He is a person who exposes any kind of information or activity that is deemed illegal, or unethical within an organization. His adversaries are often very powerful.

We suppose that he can be under a mass surveillance program, but he can not be under a targeted surveillance (for example, be followed day and night).

We also assume that the mass surveillance technologies can alert the whistleblower's supervisor if he installs Tor or other software of this type on his personal computer, they can also store and scrutinize all his communication.

{\bf The Communication channels}

There exist two communication channels, the first one is the street where people can find the secret key and the second one is the social networks were the users can find the encrypted message. The person who will be able to decrypt the cipher would be the one who has found randomly the secret key on the street.

We assume that the social networks use an end-to-end encryption ( iMessage, \dots).

{\bf The Helpers}

Anyone who walks in the street and who has at least one contact on social networks can be considered a Helper.

Regarding the messages sent abroad, we consider the Helper as a person who is ready to help dissidents from foreign countries and who has a active account on social networks.

We assume that among the 10 Helpers, there is at least one who is ready to transmit the information to journalists or activists.

{\bf The adversaries}

It can be anyone who tries to break the anonymity of the whistleblower or to prevent the diffusion of the information.

The adversaries can be:

\begin{itemize}
\item Those who are trying to prevent the diffusion of the whistleblower's information (threaten the helpers for example).
\item Those who are trying to identify the Helper.
\item Those who are trying to identify the whistleblower.
\end{itemize}

\subsubsection{Threats in our scope}

A government which stores thousands of DNA of different organisms in order to decrypt everything that passes through the networks is a plausible adversary. 
We believe that the secret services could try to monitor all messages sent through the social networks. Thus, any encrypted message, issued by a government employee could be followed (tagged). The attacker can also record messages received and sent by government employees.

A flooding attack can be tried in our scheme and a denial of service is a plausible threat.

\subsubsection{Threats not in our scope}
The adversary has no control over users' computer for example malware and spyware are not in our scope.

We consider that an attacker can not filter packets to prevent specific users from sending messages on social networks without the permission of a judge.

\subsubsection{Security-related assumptions}

We assume that the court does not per se consider organizations that relay secrets between whistleblowers and journalists as criminal. 

We suppose that the Helper will wait for a while before sending the information to journalists (the waiting time is taken randomly).

In addition, we suppose that the social networks uses an end-to-end encryption method.

We also suppose that the cryptosystem used in the end-to-end encryption possesses the property of indistinguishability.

Finally, we assume that the messages size should be the same on each social network.

\subsection{Security properties}
The first threat of this scheme will be the carelessness, in other words, when someone eats or destroys or puts in the trash the organism which should be sequenced.

{\bf Timing attack}

An attacker can try to correlate the time when an information is released with the time when a message is sent by a targeted person.
The first protection is due to the fact that the Helper, the journalist and the activist will wait for a while ( the waiting time is taken randomly) before publishing the information.

The second protection is based on the difficulty to choose one message among the others sent by a targeted person , since this latter relay all the message received and hide his message among the message to relay.

\textbf{Traffic analysis}

The encryption of a message's sending and receiving addresses (codress messages) participates in the protection against the traffic analysis \cite{hussain}.

Hence, in our scheme the message is sent without its destination and its true origin, like a ''Drunken Boat'' or a ''message in a bottle'.

Let's suppose that the Helper will wait for at least one week before sending the information to a journalist. The corrupted government will immediately try to find the Helper by verifying on its storage (obtained with the mass surveillance) all the messages sent and received by the journalist during at least one month. Let's suppose that, it will find around 10.000 encrypted messages (an end-to-end encryption is used here). Even if the indistinguishability of the cipher should protect the identity of the Helper among the ten thousand intercepted messages, however we can suppose that it would find the Helper.

The corrupted government will try to find the whistleblower by verifying on its storage all the messages sent and received by the Helper during at least one month. Let's suppose that, it will find around 10.000 encrypted messages (an end-to-end encryption is used here). The indistinguishability of the cipher should protect the identity of the whistleblower among the ten thousand intercepted messages.

However, it can try to find the path of the message with the packet header. Hence, when it follows each of these encrypted messages by trying to find the previous sender (in order to find the first sender), we affirm that it will find on each message's path, hundreds of government employee. It will also find (without knowing them), hundreds of government employee's relatives or friends. It will not find the first sender of the message since this latter can hide his message among those he should simply relay.
Hence, accusing one person will be illegal in a court, due to the fact that hundred other people can be responsible of this  disclosure.

\textbf{Denial of service}

This attack is possible, if the adversary has the power to shut down the social network or to detect and to recover immediately any abandoned organism in the street.
This attack is also possible, if the attacker can physically find and surveille all the potential whistleblowers. 
A flooding attack can be avoided if the number of message sent by a Helper is limited during a period of time.

{\bf Brute force attack}

A powerful government can create a database of the  DNA of all the organisms sold in its soil, in order to decrypt the intercepted messages. 
This attack would fail, since in addition to the DNA, the government should guess the sentence (s) taken randomly in the book and the name of the street also taken randomly.

{\bf Tagging}

If someone stops to relay a tagged message, the agents who follow that message could guess that the author of that latter is the whistleblower, however this attack is not possible in our scheme since any message is relayed in our scheme until the end of its life time.

In other words, the contact relays the encrypted message before verifying if he can decrypt it or not. 

In the same time, the attacker will find his tagged message evrywhere in the social network, which means that the monitoring of those who have received the tagged message will be difficult and unnecessary since they will receive thousand other messages which can be the real information.

{\bf Attack on the size}

The size of the messages can help an adversary to find the issuer of the message, hence, in order to fight against this attack we should have a standard size for all messages in the social network.

{\bf Physical spying }

A whistleblower who is under physical surveillance (followed day and night \dots ) should not use this method since his books can be intercepted ( where his DNA or fingerprint can peharps be found). If the whistleblower did not forget her DNA or her fingerprint in the purchased used book (containing certainly others DNA) the whistleblower can have good arguments in a court.
Instead of using a DNA Cryptography, if she writes in the book the secret key, and if the book is intercepted, the writing can lead to a graphology test in a court.

\section{Conclusion}
The basic idea of these algorithms is the fact that social networks can help to find anyone in the world in a short time.

Thus, we started  by the fact that we know someone, who knows someone, ... until someone who knows the person that we are looking for (in other words the world is interconnected). 
With the security brought by cryptographic tools, this interconnection will enable us to securely find our biological parents but also to put us immune from mass surveillance. 
Finally contrary to the ``bottle in the sea'' during the previous centuries, and which was taking a long time before reaching its destination, the speed of social networks should enable us to receive our messages in a very short time.

In the future, it could be intersting to use the authentication introduced by \cite{Brainard}, in order to introduce a voucher system in our design.


\addcontentsline{toc}{chapter}{Bibliographie}
\begin{thebibliography}{9}

\bibitem{adleman}
Adleman, Leonard M. 
\emph{Molecular computation of solutions to combinatorial problems.}  Science,  1994, vol. 266, no 5187, p. 1021-1024.

\bibitem{Brainard}
Brainard, J., Juels, A., Rivest, R. L., Szydlo, M., \& Yung, M.
\emph{Fourth-factor authentication: somebody you know.} In Proceedings of the 13th ACM conference on Computer and communications security (pp. 168-178), 2006.

\bibitem{Bruekers}
Bruekers, F., Katzenbeisser, S., Kursawe, K., \& Tuyls, P. 
\emph{Privacy-Preserving Matching of DNA Profiles.} IACR Cryptology ePrint Archive, 2008, 203.

\bibitem{Camilla}
IP, Camilla LC, Loose, Matthew, Tyson, John R., et al.
\emph {MinION Analysis and Reference Consortium: Phase 1 data release and analysis}, F1000Research, 2015, vol. 4.

\bibitem{Dingledine}
Dingledine R., Mathewson N.,  Syverson P. 
\emph{Tor: The second-generation onion router.} Naval Research Lab Washington DC, (2004).

\bibitem{ddc}
DNA Diagnostic Center
\emph{Paternity and Family Relationship Testing} http://www.dnacenter.com/science-technology/paternity-science.html

\bibitem{minidna}
EMBL-EBI,
\emph{Mini DNA sequencer tests true} http://www.ebi.ac.uk/about/news/press-releases/mini-dna-sequencer-tests-true.

\bibitem{Gearheart}
Gearheart, Christy M., Benjamin Arazi, and Eric C. Rouchka. 
\emph{DNA-based random number generation in security circuitry.} Biosystems 100.3 (2010): 208-214.

\bibitem{theguardian}
The guardian
\emph{Handheld DNA reader revolutionary and democratising, say scientists }
https://www.theguardian.com/science/2015/oct/15/handheld-dna-reader-revolutionary-and-democratising-say-scientists.

\bibitem{argentina}
The guardian
\emph{child argentina disappeared} www.theguardian.com/lifeandstyle/2014/dec/27/child-argentinas-disappeared-new-family-identity

\bibitem{france}
The guardian
\emph{France faces up to scandal of Reunion's stolen children} http://www.theguardian.com/world/2014/feb/16/france-reunion-stolen-children

\bibitem{hussain}
Hussain, AMJ Niyaz et Deepa, C.
\emph{Analysis for Traffic and Intrusion Detection.} 

\bibitem{jacob}
Jacob, Grasha. A. Murugan
\emph{DNA based cryptography: An overview and analysis.} International Journal of Emerging Sciences, 2013, vol. 3, no 1, p. 36.

\bibitem{polymorphisms}
JRank Articles, 
\emph{Polymorphisms - Strs, Vntrs, And Snps - Differences, Genetic, Sequence, and Pairs - JRank Articles}  http://medicine.jrank.org/pages/2664/Polymorphisms-STRs-VNTRs-SNPs.htmlixzz41hNQ8t00

\bibitem{kane}
Kane, Amadou Moctar.
\emph{On the use of Continued Fractions for Stream Ciphers.} In : Security and Management. 2009. p. 583-589.

\bibitem{menezes}
Menezes, Alfred J., Van Oorschot, Paul C., and Vanstone, Scott A.
\emph{Handbook of applied cryptography.} CRC press, 1996.

\bibitem{biotechlearn}
The New Zealand Biotechnology Learning Hub,
\emph{biotechlearn} http://biotechlearn.org.nz/focus\_stories/forensics/dna\_profiling

\bibitem{Babatunde}
Okunoye Babatunde, O.  
\emph{On Pseudorandom Number Generation from Programmable and Computable Biomolecules: Deoxyribonucleic (DNA) as a Novel Pseudorandom Number Generator.} 2011.

\bibitem{oxforddictionaries}
Dictionaries, Oxford.
\emph{Oxford dictionaries.} Electronic resource. Online Oxford Dictionaries - Mode of access: http://www.oxforddictionaries.com/definition/english/dna?q=DNA, Accessed March 2016.

\bibitem{Palacios}
Palacios, M. A., Benito-Pe\~na, E., Manesse, M., Mazzeo, A. D., LaFratta, C. N., Whitesides, G. M., \& Walt, D. R.
\emph{InfoBiology by printed arrays of microorganism colonies for timed and on-demand release of messages} Proceedings of the National Academy of Sciences, 108(40), 16510-16514, 2011.

\bibitem{roth}
Roth, V., G\"uldenring, B., Rieffel, E., Dietrich, S., \& Ries, L. 
\emph{A secure submission system for online whistleblowing platforms.} In Financial Cryptography and Data Security (pp. 354-361). Springer Berlin Heidelberg. 2013.

\bibitem{tornea}
Tornea, Olga. 
\emph{Contributions to DNA cryptography: applications to text and image secure transmission.} 2013. These de doctorat. Universite Nice Sophia Antipolis; Technical University of Cluj-Napoca (Roumanie).

\bibitem{tracey}
Tracey, Martin.
\emph{Short tandem repeat-based identification of individuals and parents.} Croatian medical journal, 2001, vol. 42, no 3, p. 233-238.

\bibitem{wang}
Wang, Yuhua, Wang, Guoyin, et Zhang, Huanguo. 
\emph{Random Number Generator Based on Hopfield Neural Network and SHA-2 (512).} In : Advancing Computing, Communication, Control and Management. Springer Berlin Heidelberg, 2010. p. 198-205.

\bibitem{dnaprofiling}
Wikipedia, 
\emph{DNA profiling} https://en.wikipedia.org/wiki/DNA\_profiling

\end{thebibliography}
\end{document}